\documentclass[aps,prd,superscriptaddress,twoside,twocolumn,nofootinbib,%
10pt,showpacs,floatfix]{revtex4-1}

\usepackage{amsmath,amssymb}
\usepackage{graphicx,bm}
\usepackage{ulem}
\usepackage{slashed}
\usepackage{dcolumn}
\usepackage{epsfig}
\usepackage{multirow}

\allowdisplaybreaks

\begin{document}

%%%%%%%%%%%%%%%%%%%%% Title %%%%%%%%%%%%%%%%%%%%%%%%

\title{Tetraquark mixing supported by the partial decay widths of two light-meson nonets}

%%%%%%%%%%%%%%%%%%%%%%%%%%%%%%%%%%%%%%%%%%%%%%%%

\author{Hungchong Kim}%
\email{bkhc5264@korea.ac.kr}
\affiliation{Center for Extreme Nuclear Matters, Korea University, Seoul 02841, Korea}

\author{K. S. Kim}%
\email{kyungsik@kau.ac.kr}
\affiliation{School of Liberal Arts and Science, Korea Aerospace University, Goyang, 412-791, Korea}

\date{\today}

%%%%%%%%%%%%%%%%%%%% Abstract %%%%%%%%%%%%%%%%%%%%%

\begin{abstract}

Recently, the tetraquark mixing framework has been proposed as a possible structure
for the two light-meson nonets in the $J^P=0^+$ channel, the light nonet composed of $a_0(980)$, $K_0^*(700)$, $f_0(500)$, $f_0(980)$,
and the heavy nonet of $a_0(1450)$, $K_0^*(1430)$, $f_0(1370)$, $f_0(1500)$.
Among various signatures, we report in this work that their partial decay widths collected from various experimental data in Particle Data Group (PDG)
can support this mixing scheme also.
In particular, we demonstrate that the couplings of the light nonet to two pseudoscalar mesons estimated from the
partial widths are consistently stronger than those of the heavy nonet.
This consistent feature agrees qualitatively well with the predictions from the tetraquark mixing framework
and, therefore, provides supporting evidence for the tetraquark mixing.

\end{abstract}

%\pacs{
%14.40.Rt,	% Exotic mesons
%14.40.Be,	% Light mesons (S=C=B=0)
%14.40.Df,   % Strange mesons (|S|>0, C=B=0)
%11.30.Hv    % symmetries
%}

\maketitle

\section{Introduction}

In the meson spectrum with the spin-parity of $J^P=0^+$~\cite{PDG22}, there are two sets of resonances
which seem to form a SU(3) flavor nonet ($\bm{9}_f$) separately (see Table~\ref{two nonets}). One set is composed of
$a_0 (980)$, $K_0^* (700)$, $f_0 (500)$, $f_0 (980)$, and we refer them as the light nonet.
The other set is the heavy nonet composed of $a_0 (1450)$, $K_0^* (1430)$, $f_0 (1370)$, $f_0 (1500)$.
%%%%%%%%%%%%%%%%%%%%%%%%%%%%%%%%%%%%%%%%%%%%%%%%%%%%%%%%%%%%%%%%%%%%%%%
Their isospin composition ($I=0,1/2,1$) is the same as the lowest-lying pseudoscalar mesons ($\pi$,$K$,$\eta$,$\eta^\prime$)
and the vector mesons ($\rho$,$K^*$,$\omega$,$\phi$), so it might be reasonable to assume that the two nonets separately form
a flavor nonet. This assumption is also supported by the fact
that the two nonets approximately satisfy the Gell-Mann--Okubo mass relation~\cite{Kim:2018zob}.
Even though the two nonets have the same isospin structure and spin-parity, they are highly separated in mass around 500 MeV or more.
It would be interesting to look for a specific framework to describe the two nonets.
%%%%%%%%%%%%%%%%%%%%%%%%%%%%%%%%%%%%%%%%%%%%%%%%%%%%%%%%%%%%%%%%%%%%%%%

Indeed, there have been various approaches to describing the two nonets.
They include hadronic molecular picture for
$f_0(980)$, $a_0(980)$~\cite{Janssen:1994wn,Branz:2007xp,Branz:2008ha,Dudek:2016cru},
a two-quark picture with hadronic intermediate states~\cite{Tornqvist:1995kr, vanBeveren:1986ea},
tetraquarks mixed with a glueball~\cite{Maiani:2006rq}, the mixing model of the $P$-wave $q\bar{q}$ with the four-quark
$qq\bar{q}\bar{q}$ scalar nonet~\cite{Black:1998wt,Black:1999yz}, and the tetraquarks including instantons~\cite{Dorokhov:1993nw}.

Maybe the most famous one is the diquark-antidiquark model~\cite{Jaffe77a,Jaffe77b,Jaffe04} to describe the light nonet.
In this approach, the light nonet is regarded as a tetraquark nonet ($\bm{9}_f$) constructed from diquark and antidiquark.
The spin-0 diquark, in the color and flavor structures of ($\bar{\bm{3}}_c, \bar{\bm{3}}_f$),
is adopted in this construction as it is most compact among all possible diquarks when the binding is calculated from
the color-spin interaction.
This tetraquark model is strongly supported by the inverted mass ordering
of $M[a_0(980)] > M[K_0^*(700)] > M[f_0(500)]$ as this ordering
cannot be established in a two-quark picture with the orbital angular momentum $\ell=1$.
We denote this tetraquark type by $|000\rangle$ where the first zero is the tetraquark spin,
the second the diquark spin and the third the antidiquark spin.
One problem with this approach could be that the light nonet members, whose masses are being below 1 GeV, are too light to be tetraquarks.

%%%%%%%%%%%  Table 1,  two nonets in PDG

\begin{table}%[t]
\centering
\begin{tabular}{c|c|c|c||c|c|c}  \hline\hline
$I$ &  LN  & Mass& $\Gamma_{tot}$&  HN  & Mass & $\Gamma_{tot}$ \\
\hline
$ 1 $ & $a_0 (980)$  & 980 &  50-100  & $a_0 (1450)$ & 1474 &  265\\
 $1/2$ & $K_0^* (700)$ & 845 & 468 & $K_0^* (1430)$ & 1425 & 270  \\
$ 0 $ & $f_0 (500)$  & 400-800 &  100-800 & $f_0 (1370)$ & 1200-1500 &  200-500  \\
$ 0 $ & $f_0 (980)$  & 990 &  10-100 & $f_0 (1500)$ & 1506 &  112 \\
\hline
\end{tabular}
\caption{The light nonet (LN) and heavy nonet (HN) in the $J^P=0^+$ channel collected from PDG~\cite{PDG22} are listed here
along with the isospin, mass, the total width. The masses of $f_0 (500)$, $K_0^* (700)$, which are now named as
the Breit-Wigner mass, have been changed substantially
from their values in the 2016 version of PDG~\cite{PDG16}. Mass and width are given in MeV unit.}
\label{two nonets}
\end{table}
%%%%%%%%%%%%%%%%%%%%%%%%%%%%%%%%%%%%%%%%%%%%%%%%%%%%%%%%%%%%%%%%%%%%%%%%%%%%%%%%%%%%%%%%%%%%%%%%%%%%%%%%%%

The heavy nonet members, although their splitting is marginal, have similar mass ordering
like $M[a_0(1450)] > M[K_0^*(1430)] > M[f_0(1370)]$ if one takes the central masses from their experimentally known range in Table~\ref{two nonets}.
So one may attempt to describe the heavy nonet as tetraquarks also.
To do this, we have advocated the second tetraquark type with different diquark
configuration~\cite{Kim:2016dfq,Kim:2017yur,Kim:2017yvd,Kim:2018zob,Lee:2019bwi,Kim:2019mbc}.
Specifically, the spin-1 diquark, in the color and flavor structures of
($\bm{6}_c, \bar{\bm{3}}_f$), is used to construct a spin-0 tetraquark nonet ($\bm{9}_f$).
We denote this second tetraquark by $|011\rangle$ to represent the spin-0 tetraquark
formed by spin-1 diquark and spin-1 antidiquark.
Note, the spin-1 diquark is less compact than the spin-0 diquark above but the second tetraquark, $|011\rangle$,
constructed from this spin-1 diquark is found to be more compact than the first tetraquark type, $|000\rangle$,
if the binding is calculated from the color-spin interaction applied for all the pairs of the constituents.
Our standpoint is that the second tetraquark must be considered
along with the first tetraquark when one deals with tetraquarks.
One remaining issue is how to match the two tetraquark types to the two nonets in PDG.

By construction, the two tetraquark types, $|000\rangle$ and $|011\rangle$, have the same flavor structure ($\bm{9}_f$), but
they have different configurations in color and spin. This means, for each member in the flavor nonet, there are two configurations
distinguished by the color and spin structure. The important observation is that these two configurations mix
through the color-spin interaction,
$V_{CS} \propto\sum_{i < j}  \frac{\lambda_i \cdot \lambda_j J_i\cdot J_j}{m_i^{} m_j^{}}$~~\footnote{Here, $\lambda_i$ denotes the Gell-Mann matrix for color and
$J_i$ the spin, $m_i$ the constituent quark mass.}.
The mixing term like $\langle 000| V_{CS}|011\rangle$ is found to be large~\cite{Kim:2016dfq} so
the expectation value of the color-spin interaction,
$\langle V_{CS} \rangle$, normally known as the hyperfine mass, actually forms a $2\times 2$ matrix when it is
evaluated in the basis, $|000\rangle$ and $|011\rangle$.
In other words, $|000\rangle$, $|011\rangle$ are not the eigenstates of the color-spin interaction, and, therefore, can not
represent the physical resonances.  Instead, the eigenstates that diagonalize the $2\times 2$ matrix can be identified
as physical resonances hopefully realized by the light and heavy nonets in PDG~\cite{PDG22}.

To test this scenario, the steps that we took in Refs.~\cite{Kim:2016dfq,Kim:2017yur,Kim:2017yvd,Kim:2018zob} are as follows.
For each flavor member, we calculate the $2\times 2$ hyperfine matrix for $V_{CS}$ in the basis of $|000\rangle$ and $|011\rangle$.
We then diagonalize the matrix to get the physical hyperfine masses (as given in Table~\ref{parameters}) and the corresponding
eigenstates,
\begin{eqnarray}
|\text{Heavy~nonet} \rangle &=& -\alpha | 000 \rangle + \beta |011 \rangle \label{heavy}\ ,\\
|\text{Light~nonet} \rangle~ &=&~~\beta | 000 \rangle + \alpha |011 \rangle \label{light}\ ,
\end{eqnarray}
which are identified as the corresponding members of the heavy and light nonets in PDG.
The mixing parameters $\alpha,\beta$ are also fixed by the diagonalization.
They are calculated separately for each member of the flavor nonet
but, as shown in Table~\ref{parameters}, their values almost do not depend on the isospin channel.
This means that the left-hand sides of Eqs.~(\ref{heavy}),(\ref{light}), if collected for all the
nonet members, also form a flavor nonet separately just as $|000\rangle$, $|011\rangle$.

It is clear from Eqs.~(\ref{heavy}),(\ref{light}) that $\alpha,\beta$ represent the probability amplitude to find
the physical states in the configurations, $|000\rangle$, $|011\rangle$.
Since $\alpha > \beta$, the light nonet members, $a_0 (980)$, $K_0^* (700)$, $f_0 (500)$, $f_0 (980)$,
represented by Eq.~(\ref{light}), have more probability to stay
in the $|011 \rangle$ configuration rather than in the $|000 \rangle$. This feature is quite different
from the normal assumption that the light nonet is in the $|000 \rangle$ configuration.
This model, where the two nonets are represented by Eqs.~(\ref{heavy}),(\ref{light}), is the {\it tetraquark mixing model}
that we have proposed and tested in various occasions
in Refs.~\cite{Kim:2016dfq,Kim:2017yur,Kim:2017yvd,Kim:2018zob,Lee:2019bwi,Kim:2019mbc}.

This tetraquark mixing model has some successful features that
match well with the experimental facts.
First, the huge mass gap, around $\Delta M\approx 500$ MeV, between the light and heavy nonets, is explained very well
by the hyperfine mass splitting, $\Delta \langle V_{CS} \rangle$, (see Table~\ref{parameters}),
satisfying the mass splitting formula, $\Delta M\approx \Delta \langle V_{CS} \rangle$~\cite{Kim:2016tys,Kim:2014ywa,Kim:2016dfq}.
Second, the mixing model generates the negatively huge hyperfine masses for the light nonet,
around $\langle V_{CS} \rangle\approx -500$ MeV,
which can explain why the light nonet members, in spite of being tetraquarks, have such small masses below
1 GeV~\cite{Kim:2016dfq,Kim:2017yvd,Kim:2018zob,Kim:2019mbc}.
On the other hand,  the small hyperfine masses of the heavy nonet, around $\langle V_{CS} \rangle\approx -20$ MeV,
which are also driven mainly by the mixing framework,
can explain why the heavy nonet members have the masses not far from $4m_q$, four times of the constituent quark mass.
In addition, the marginal mass ordering seen in the heavy nonet can be {\it partially} explained by the small splitting among the hyperfine masses
of the heavy nonet~\cite{Kim:2019mbc}.

%%%%%%%%%%%  Table 2: mixing parameters and hyperfine masses
\begin{table}[t]
\centering
%\tbl{Hyperfine masses and the mixing parameters for the two nonets~\cite{2019mbc}.}
\begin{tabular}{c|c|c||c|c||c|c}\hline\hline
$I$ &  LN & $\langle V_{CS} \rangle$ & HN & $\langle V_{CS} \rangle$  & $\alpha$ & $\beta$ \\
\hline
%[-2.8mm]
$1$ & $a_0(980)$    & $-488.5$ &$a_0(1450)$    & $-16.8$ & 0.8167     & 0.5770    \\
$1/2$ & $K_0^*(700)$  & $-592.7$ &$K_0^*(1430)$  & $-26.9$ & 0.8130     & 0.5822    \\
$0$& $f_0(500)$    & $-667.5$ &$f_0(1370)$    & $-29.2$ & 0.8136     & 0.5814    \\
 \hline
$0$& $f_0(980)$  & $-535.1$ &$f_0(1500)$     & $-20.1$ & 0.8157     & 0.5784    \\
\hline
\end{tabular}
\caption{Here are the physical hyperfine masses, $\langle V_{CS} \rangle$,
and the mixing parameters, $\alpha$, $\beta$, for the two nonets
calculated separately in each isospin channel~\cite{Kim:2019mbc}. Note,
$\alpha$, $\beta$, are almost the same regardless of the isospin channel.
See the text for the physical meaning of $\alpha$, $\beta$.
$\langle V_{CS} \rangle$ is given in MeV unit.}
\label{parameters}
\end{table}
%%%%%%%%%%%%%%%%%%%%%%%%%%%%%%%%%%%%%%%%%%%%%%%%%%%%%%%%%%%%%%%%%%%%%%%%%%%%%%%%%

The most striking feature of the tetraquark mixing model is the
coupling strength that governs the two nonets decaying into two pseudoscalar
mesons.  The two tetraquark types, $|000\rangle$ and $|011\rangle$, can decay into two pseudoscalar mesons
through the fall-apart mechanism~\cite{Jaffe77b} where the decay proceeds through a recombination of quark and antiquark into
the component of two color-singlet, $\bm{1}_c\otimes\bm{1}_c$,
obtained from the rearrangement, $(qq)(\bar{q}\bar{q})\rightarrow (q\bar{q})(q\bar{q})$.
Due to the relative sign difference between Eqs.~(\ref{heavy}), (\ref{light}),
the coupling strength for this decay is enhanced for the light nonet and suppressed for the heavy nonet.

The purpose of this paper is to test the last feature of the tetraquark mixing model for the full members of the two nonets.
Note, this feature has
been investigated already but only for the isovector members, $a_0(980)$, $a_0(1450)$, in Ref.~\cite{Kim:2017yur},
and we found that the couplings are indeed enhanced for the $a_0(980)$,
while suppressed for the $a_0(1450)$.
At that time, this type of investigation was not extended to the other
members because, first of all, some experimental data were not available
and, theoretically, the tetraquark mixing model was not fully developed to include the isoscalar resonances
$f_0(500)$, $f_0(980)$, $f_0(1370)$, $f_0(1500)$.
With more experimental data available and theoretical progresses~\cite{Kim:2017yvd,Kim:2018zob,Kim:2019mbc},
we are now able to extend this interesting test to the other resonances in the two nonets.
In doing so, we hope that this universal trend in the coupling strengths can be established so that the tetraquark mixing framework
can be regarded as a relevant structure for the two nonets in PDG.

This paper is organized as follows.
In Sec.~\ref{gamma}, the experimental partial decay widths of the two nonets are extracted from current data in PDG~\cite{PDG22}.
These partial widths are then used in Sec.~\ref{implication} to motivate that the coupling strengths of the light nonet to
two pseudoscalar mesons are much stronger than those of the heavy nonets.
In Sec.~\ref{couplings from TMM}, we discuss
how the tetraquark mixing model can explain consistently this interesting phenomenology on the couplings.
We then calculate the theoretical partial widths using the couplings from the tetraquark mixing model
and compare them with the corresponding experimental partial widths in Sec.~\ref{comparison}.
We summarize in Sec.~\ref{sec:summary}.

\section{Experimental Partial Widths}
\label{gamma}

The adequate quantities in testing the enhancement and suppression of the couplings
would be the partial decay widths that are experimentally accessible both from the light nonet and the heavy nonet.
The partial decay width is basically a function of the coupling strength and
the kinematical factors that depend on the mass difference between the initial and final states of a decay.
It means, only after the experimental partial widths at hand,
we might be able to discuss how the coupling strengths should depend on the two nonets.
For this purpose, we extract experimental partial widths from the two nonets based on the PDG data~\cite{PDG22}.
Since most partial widths are not explicitly stated in PDG,
%Still a challenge is that
%most of the partial widths are not clearly known in PDG and, thus,
it may be useful to discuss how we extract the experimental partial widths
from the current PDG so that
future readers can utilize this information for further improvements as the PDG update is progressing.

For $a_0(980)$, the partial decay width, $\Gamma[a_0(980)\rightarrow \pi \eta] \approx 60$ MeV, was taken following the statement
in PDG, ``Peak width in $\eta \pi$ is about 60 MeV but decay width can be much larger''.
The partial width $\Gamma[a_0(980)\rightarrow K \bar{K}]\approx 10.6$ MeV is obtained
from the experimental ratio of $\Gamma[a_0(980)\rightarrow K \bar{K}]/\Gamma[a_0(980)\rightarrow \pi \eta]\approx 0.177$~\cite{PDG22}.
Note that this ratio was 0.183 in the 2016 version of PDG~\cite{PDG16}.

For $a_0(1450)$, most branching ratios for its decays are rather clearly reported in PDG~\cite{PDG22}
except for the one poorly known ratio~\cite{Anisovich:2001jb},
%\begin{eqnarray}
$0\leq R \equiv \frac{\Gamma[a_0(1450)\rightarrow a_0(980) \pi\pi]}{\Gamma [a_0(1450)\rightarrow \pi \eta]} \leq 4.3$.
%\end{eqnarray}
PDG quotes that this measurement has not been used in the analysis $a_0(1450)$ but, since its upper limit is somewhat large,
we take this range of $R$ as an additional source of the uncertainty in the extraction of the partial widths.
Using the five decay modes reported in PDG and equating their sum to the full width of 265 MeV, we extract
the partial widths of our interest in the two limiting cases, $R=0$ and $R=4.3$.
We find that $\Gamma[a_0(1450)\rightarrow \pi \eta] \approx 20.5$ MeV, $\Gamma[a_0(1450)\rightarrow K \bar{K}] \approx 18.0$ MeV
when $R=0$, and when $R=4.3$, we get
$\Gamma[a_0(1450)\rightarrow \pi \eta] \approx 15.4$ MeV, $\Gamma[a_0(1450)\rightarrow K \bar{K}] \approx 13.5$ MeV as listed in Table~\ref{partial width}.
In this estimate, we have neglected the additional partial width of $a_0(1450)\rightarrow \gamma \gamma$ in PDG as it is an order of 10 keV at most.

For $K_0^*(700)$, PDG reported that $K_0^*(700)$ decays to $\pi K$ with branching fraction of 100\%. So the $\pi K$ partial width
is taken to be the total decay width of $K_0^*(700)$, $468$ MeV.
This experimental partial width is new in PDG, not reported in the old versions of PDG published before the year 2020.
Now, with this new information at hand, we are able to test the tetraquark mixing model also through the decays from $K_0^*(700)$,
$K_0^*(1430)$.
For $K_0^*(1430)$, the $\pi K$ partial width is extracted from its branching ratio, $0.93$, and the total width $270$ MeV,
i.e., $\Gamma[K_0^*(1430)\rightarrow \pi K]=270\times 0.93=251.1$ MeV.

For $f_0(980)$,  $\Gamma[f_0(980)\rightarrow \pi \pi] \approx 50$ MeV was taken following the statement
in PDG, ``Peak width in $\pi \pi$ is about 50 MeV but decay width can be much larger''.
The partial width for $f_0(980)\rightarrow K \bar{K}$ is scarcely known in PDG~\cite{PDG22}.
Instead, there are several experimental values available for the branching ratio of $r\equiv\Gamma(\pi\pi)/[\Gamma(\pi\pi)+\Gamma(K\bar{K})]$
but PDG quotes that these measurements have not been used in the analysis of $f_0(980)$.
These measurements lead to the broad range, $0.52 \le r \le 0.84$~\cite{BaBar:2006hyf,BES:2005iaq,Anisovich:2001ay,CERN:1979etn,Cason:1978zp,Wetzel:1976gw}.
Using this broad range and $\Gamma[f_0(980)\rightarrow \pi \pi] \approx 50$ MeV,
we find the partial width for $\Gamma[f_0(980)\rightarrow K\bar{K}]$ with huge uncertainty like $9.52 \text{--} 46.2$ MeV.
For $f_0(1500)$, we extract the partial widths from the total width of $\Gamma_{tot}=112$ MeV
and the branching fractions reported in PDG, ${\cal B}(f_0(1500)\rightarrow \pi \pi)\approx 0.34$,
${\cal B}(f_0(1500)\rightarrow K \bar{K})\approx 0.085$.  This procedure yields $\Gamma[f_0(1500)\rightarrow \pi \pi]\approx 38.1$ MeV,
$\Gamma[f_0(1500)\rightarrow K \bar{K}]\approx 9.5$ MeV

%%%%%%%%%%%  Table 3,  partial widths from PDG

\begin{table}%[t]
\centering
\begin{tabular}{l|c|l|c}  \hline\hline
\multicolumn{2}{c|}{Light nonet} & \multicolumn{2}{c}{Heavy nonet} \\
\hline
     Decay mode &  $\Gamma_{exp}$(MeV)  & Decay mode & $\Gamma_{exp}$(MeV) \\
\hline
$~a_0(980)\rightarrow \pi \eta$  &$60$ & $~a_0(1450)\rightarrow \pi \eta$&  $15.4\text{--}20.5$  \\
%                                         &      &                                           &  ~$~15.8$ (set 2) \\
%\hline
$~a_0(980)\rightarrow K\bar{K}$  &10.6 & $~a_0(1450)\rightarrow K\bar{K}$ &  $13.5\text{--}18.0$ \\
%                                         &      &                                          &  ~$~13.5$ (set 2) \\
\hline
$~K_0^*(700)\rightarrow \pi K$ & $468$  & $~K_0^*(1430)\rightarrow \pi K$ &  $251.1$   \\
\hline
%                                         &      &                                          &  \\
$~f_0(500)\rightarrow \pi \pi$ & NA  & $~f_0(1370)\rightarrow \pi \pi$ &  NA   \\
\hline
$~f_0(980)\rightarrow \pi \pi$ & $50$  & $~f_0(1500)\rightarrow \pi \pi$ &  $38.1$   \\
$~f_0(980)\rightarrow K\bar{K}$ & $9.5\text{--}46.2$  & $~f_0(1500)\rightarrow K\bar{K}$ & $9.5$  \\

\hline
\end{tabular}
\caption{Partial decay modes and the corresponding widths collected from PDG~\cite{PDG22} are listed here for the light and heavy nonets.
The partial widths from $f_0(500)$, $f_0(1370)$ are not available (NA) at the moment due to the lack of experimental data.
Since our purpose is to compare the common decay modes in the two nonets,
the other decay modes not present for
%both are
both nonets have been omitted from this table.}
\label{partial width}
\end{table}
%%%%%%%%%%%%%%%%%%%%%%%%%%%%%%%%%%%%%%%%%%%%%%%%%%%%%%%%%%%%%%%%%%%%%%%%%%%%%%%%%%%%%%%%%%%%%%%%%%%%%%%%%%

\section{Coupling strengths from the experimental partial widths}
\label{implication}

Now it is possible to extract interesting characteristics
from the experimental partial widths presented in Table~\ref{partial width}.
What we want to demonstrate is that the current partial width data
actually imply that the coupling strengths of the light nonet to two pseudoscalar mesons
are consistently enhanced relative to the corresponding coupling strengths of the heavy nonet.

To illustrate this, first we note that most partial widths from the light nonet are larger than
the corresponding widths from the heavy nonet. In particular,
Table~\ref{partial width} shows that
\begin{eqnarray}
&&\Gamma_{exp}[a_0(980)\rightarrow \pi \eta] > \Gamma_{exp}[a_0(1450)\rightarrow \pi \eta]\label{gam1}\ ,\\
&&\Gamma_{exp}[K_0^*(700)\rightarrow K \bar{K}] > \Gamma_{exp}[K_0^*(1430)\rightarrow K \bar{K}]\label{gam2}\ ,\\
&&\Gamma_{exp}[f_0(980)\rightarrow \pi \pi] > \Gamma_{exp}[f_0(1500)\rightarrow \pi \pi]\label{gam3}\ .
\end{eqnarray}
Also, it is mostly like that
\begin{eqnarray}
\Gamma_{exp}[f_0(980)\rightarrow K \bar{K}] > \Gamma_{exp}[f_0(1500)\rightarrow K \bar{K}]\label{gam4}\ ,
\end{eqnarray}
if we exclude the region around the lower limit, $\Gamma_{exp}[f_0(980)\rightarrow K \bar{K}] \sim 9.5$.
These experimental inequalities seem strange if one looks at them purely from the kinematical point of view.
Because of the huge mass difference, around $500$ MeV, the heavy nonet
has much more phase space available for its decays than the corresponding light nonet.
A naive expectation would be then that the partial widths from the heavy nonet
are larger than those of the light nonet but what we see from the inequalities above is opposite to this expectation.

This naive expectation can be estimated in fact by calculating the partial widths directly using the effective
Lagrangian for the light (heavy) nonet and two-pseudoscalar mesons
involving derivatives~\cite{Black:1999yz}. The effective Lagrangian must have the same form for the light and heavy nonets
since both nonets separately form a nonet in SU(3)$_f$.
For a process where a nonet member with mass $M$ decays into two pseudoscalars with masses, $m_1$,$m_2$,
the partial width can be calculated by
\begin{eqnarray}
\Gamma [M\rightarrow m_1,m_2] &=& G^2 \frac{p}{32\pi M^2} (M^2-m_1^2-m_2^2)^2\nonumber \\
&\equiv& G^2 \Gamma_{kin}\ ,
\label{decay width}
\end{eqnarray}
where $p$ is the momentum of the decay products in the center of mass frame.
The constant, $G$, is the coupling strength of the effective Lagrangian,
$G~\partial_\mu \psi_{m_1} \partial^\mu \psi_{m_2} \psi_M$, in this channel.
The kinematical factors like $p$ and $M^2-m_1^2-m_2^2$, in Eq.~(\ref{decay width})
increase as the mass gap between the initial and final states in the decay increases.
To discuss the decay width driven solely by such kinematical factors, we have also defined the
``kinematical decay width'', $\Gamma_{kin}$, as in Eq.~(\ref{decay width}).

To make a crude estimation, we simply take the masses in Table~\ref{two nonets},
calculate $\Gamma_{kin}$ for $a_0(980)\rightarrow \pi \eta$, $a_0(1450)\rightarrow \pi \eta$,
and take a ratio of them to estimate the enhancement purely from the kinematical widths,
\begin{eqnarray}
\frac{\Gamma_{kin}[a_0(1450)\rightarrow \pi \eta]}{\Gamma_{kin}[a_0(980)\rightarrow \pi \eta]}\approx 1.68\ .
\label{kin ratio}
\end{eqnarray}
Thus, as we have mentioned already, this kinematical consideration alone
leads to the same conclusion agreeing with
the naive expectation that
the $a_0(1450)\rightarrow \pi \eta$ width is much larger than the $a_0(980)\rightarrow \pi \eta$ width.
But, the experimental ratio is opposite to Eq.~(\ref{kin ratio}) as one can see from the inequality in Eq.~(\ref{gam1}).
The actual experimental ratio estimated from Table~\ref{partial width} is
\begin{eqnarray}
\frac{\Gamma_{exp}[a_0(1450)\rightarrow \pi \eta]}{\Gamma_{exp}[a_0(980)\rightarrow \pi \eta]}\approx 0.26 \text{--} 0.34\ ,
\label{a0 width}
\end{eqnarray}
so, in reality, the $a_0(980)$ partial width is much larger than the $a_0(1450)$ partial width.

It is clear from Eq.~(\ref{decay width}) that only resolution for this mismatch
between Eqs.~(\ref{kin ratio}),(\ref{a0 width}) is to have different coupling strength, $G$,
for the two resonances.  More precisely, the coupling strength, $G$, of $a_0(980)$ must be strongly enhanced
relative to the coupling of $a_0(1450)$ in order to overcome the kinematical enhancement of Eq.~(\ref{kin ratio})
and hopefully achieve the consistency with the experimental ratio of Eq.~(\ref{a0 width}).
A similar discussion can be applied to the other three inequalities given in Eqs.~(\ref{gam2}),(\ref{gam3}),(\ref{gam4})
and the same conclusion as above can be reached, i.e., the coupling strengths of the light nonet
must be enhanced relative to those of the heavy nonet.

One exceptional case that looks different is the isovector case where the light nonet width,
$\Gamma_{exp}[a_0(980)\rightarrow K \bar{K}]\approx 10.6$ MeV, is less
than the heavy nonet width,  $\Gamma_{exp}[a_0(1450)\rightarrow K \bar{K}]\approx 13.5\text{--}18$ MeV.
In this case, however, the $a_0(980)$ mass lies just on or below the $K\bar{K}$ threshold.
Its decay, $a_0(980)\rightarrow K \bar{K}$, is mostly prohibited by the kinematical cutoff and rarely occurs
through the upper tail of the $a_0(980)$ total width.
If this kinematical cutoff is taken into account,
%one cannot say that the present experimental width of $10.6$ MeV is small
%when it is compared to heavy nonet width of $13.5\text{--}18$ MeV.
%Instead, the present width actually shows
%how strongly the coupling strength is enhanced in comparison with its counterpart in the heavy nonet (see more on this in Sec.~\ref{comparison}).
the present experimental width of $10.6$ MeV is not so small when this is compared to the heavy nonet width of $13.5\text{--}18$ MeV.
Instead, the present width shows how strongly the coupling strength of the light nonet is
enhanced compared to that of the heavy nonet (see more on this in Sec.~\ref{comparison}).
The similar analysis can be applied to the decay, $f_0(980)\rightarrow K \bar{K}$, in comparison with $f_0(1500)\rightarrow K \bar{K}$,
especially when the experimental partial width of $f_0(980)\rightarrow K \bar{K}$ is around the lower limit of $9.5$.

The last decay channels to discuss are $f_0(500)\rightarrow \pi\pi$, $f_0(1370)\rightarrow \pi\pi$
whose partial widths currently are not available from PDG.
The $f_0(500)$, also known as ``$\sigma$'' for long time,  is famous for its broad width.  Its total width currently
reported in PDG is $\Gamma_{tot}=100 \text{--} 800$ MeV,
which is expected to be saturated mostly by the partial mode of $f_0(500)\rightarrow \pi\pi$ as the other decay channels
like $K\bar{K}$ etc are kinematically suppressed.
On the other hand, the heavy nonet member, $f_0(1370)$, has the total width $\Gamma_{tot}=200 \text{--} 500$ MeV
and this total width should be divided into various partial widths that include not only the $\pi\pi$ mode but also the others like $4\pi$, $\eta\eta$, $K\bar{K}$, etc.
Even though specific numbers for the widths are not available at the moment, it is expected that
$\Gamma_{exp} [f_0(500)\rightarrow \pi\pi] > \Gamma_{exp} [f_0(1370)\rightarrow \pi\pi]$.
Therefore, the above conclusion on the coupling strengths may also hold for $f_0(500)\rightarrow \pi\pi$, $f_0(1370)\rightarrow \pi\pi$.

\section{Coupling strengths from the tetraquark mixing model}
\label{couplings from TMM}

In the previous section, we have established a universal trend that
the coupling strengths of the light nonet to two pseudoscalar mesons
are enhanced while the strengths of the heavy nonet are suppressed,
\begin{eqnarray}
|G|(\text{light nonet}) \gg |G|(\text{heavy nonet})\ .
\label{coupling ordering}
\end{eqnarray}
This conclusion was based on the analysis of the experimental partial widths in Table~\ref{partial width}.
Then, it is anticipated to have a certain dynamics that can explain this peculiar inequality in the coupling strengths.
As mentioned already,
the tetraquark mixing model~\cite{Kim:2016dfq,Kim:2017yur,Kim:2017yvd,Kim:2018zob,Lee:2019bwi,Kim:2019mbc}
can provide such an inequality between the two set of the couplings, and, therefore, it gives a reasonable theoretical
support for the conclusion above, Eq.~(\ref{coupling ordering}).
The similar discussion can be found also in our earlier publications also~\cite{Kim:2017yur,Kim:2017yvd,Kim:2019mbc}
but we repeat the discussion again in order to make this paper self-contained.

Tetraquarks can decay into two pseudoscalar mesons by fall-apart mechanism~\cite{Jaffe77b} where
quarks and antiquarks in a tetraquark labeled as $q_1 q_2 \bar{q}_3\bar{q}_4$
are recombined into $q_1\bar{q}_3$ and $q_2\bar{q}_4$, and simply fall apart into two mesons.
This decay is possible because the tetraquark in this recombination
has a component with two color-singlets
%~\footnote{This component, if viewed in spin and flavor, is further divided into two types of meson states,
%pseudoscalar-pseudoscalar and vector-vector~\cite{Kim:2018zob}.}
in addition to the hidden color component forming a color-singlet totally.
This recombination in color is schematically represented by
\begin{eqnarray}
q_1 q_2 \bar{q}_3\bar{q}_4 \sim (\bm{1}_c)_{13}\otimes (\bm{1}_c)_{24}  +
(\bm{8}_c)_{13}\otimes (\bm{8}_c)_{24} \label{recomb}\ .
\end{eqnarray}

In the tetraquarks represented by the mixing formulas, Eqs.~(\ref{heavy}),(\ref{light}),
the two tetraquark types, $|000\rangle$ and $|011\rangle$, separately, have the fall-apart modes
into two pseudoscalar mesons in the recombination like Eq.~(\ref{recomb}).
In this process,
there are accompanying numerical factors associated with flavor recombination,
spin recombination, and color recombination.
Since $|000\rangle$ and $|011\rangle$ in Eqs.~(\ref{heavy}),(\ref{light}) form a flavor nonet separately,
the numerical factors from the flavor recombination must be the same for both.
But the numerical factors from the spin and color recombination are different because
$|000\rangle$, $|011\rangle$ differ by the spin configuration and color configuration.
Specifically, $|000\rangle$ is the spin-0 tetraquark constructed from spin-0 diquark in $\bar{\bm{3}}_c$
and spin-0 antidiquark in $\bm{3}_c$, while
$|011\rangle$ is the spin-0 tetraquark constructed from spin-1 diquark in $\bm{6}_c$
and spin-1 antidiquark in $\bar{\bm{6}}_c$.
In each isospin channel, the two numerical factors partially cancel in Eq.~(\ref{heavy}) to make the heavy nonet
but they add up in Eq.~(\ref{light}) to make the light nonet.

To make this argument more clear, we write down, for example, the wave functions for the isovector resonances, $a_0(980)$, $a_0(1450)$,
in this recombined form as
\begin{eqnarray}
&&|a_0^+(980)\rangle =G_1 |\pi^+ \eta\rangle + G_2 |K^+\bar{K}^0 \rangle +\cdot\cdot\cdot\label{isol}\ ,\\
&&|a_0^+(1450)\rangle =G_1^\prime |\pi^+ \eta\rangle + G_2^\prime |K^+\bar{K}^0 \rangle +\cdot\cdot\cdot\label{isoh}\ ,
\end{eqnarray}
where ``$\cdot\cdot\cdot$'' denotes other two-meson states as well as the hidden color component.
The recombination process mentioned above leads to the coefficients as~\cite{Kim:2017yur}
\begin{eqnarray}
&&G_1=~\frac{\beta}{3\sqrt{2}}+\frac{\alpha}{\sqrt{3}}\:~; G_2=~\frac{\beta}{2\sqrt{3}}+\frac{\alpha}{\sqrt{2}}\ ,\\
&&G_1^\prime =-\frac{\alpha}{3\sqrt{2}}+\frac{\beta}{\sqrt{3}}\:; G_2^\prime =-\frac{\alpha}{2\sqrt{3}}+\frac{\beta}{\sqrt{2}}\ .
\end{eqnarray}
Through the relative signs in these coefficients, it is clearly exhibited that the adding and the partial cancelation of the two tetraquark types, $|000\rangle$, $|011\rangle$,
occur in building the light and heavy nonet members, $a_0^+(980)$, $a_0^+(1450)$.
Alternatively, from Eqs.~(\ref{isol}),(\ref{isoh}), one can see that these coefficients can be also defined as
\begin{eqnarray}
&&G_1=\langle \pi^+\eta|a_0^+(980)\rangle\:;~G_2=\langle K^+\bar{K}^0|a_0^+(980)\rangle\ ,\\
&&G_1^\prime=\langle \pi^+\eta|a_0^+(1450)\rangle\:; G_2^\prime=\langle K^+\bar{K}^0|a_0^+(1450)\rangle .
\end{eqnarray}
indicating that these coefficients in fact represent the coupling strengths up to a common constant
that should appear in the effective Lagrangian to fix the physical dimension.

One can obtain the coupling strengths of other nonet members to two-meson states similarly.
Further discussion can be found in Refs.~\cite{Kim:2017yur,Kim:2017yvd} and here we simply present the coupling strengths
for the channels that can be used for comparison with the experimental partial widths in Table~\ref{partial width},
\begin{widetext}
\begin{eqnarray}
\begin{array}{l|l||l|l}
\text{Light nonet mode} & \text{Coupling strength} (G) & \text{Heavy nonet mode} & \text{Coupling strength} (G^\prime) \\
\hline
%\\[-2.8mm]
a_0^+(980) \rightarrow \pi^+ \eta\  &~~\frac{\beta}{3\sqrt{2}}+\frac{\alpha}{\sqrt{3}}=0.6076 &a_0^+(1450) \rightarrow \pi^+ \eta & -\frac{\alpha}{3\sqrt{2}}+\frac{\beta}{\sqrt{3}}=0.1406 \\[1mm]
a_0^+(980) \rightarrow K^+\bar{K}^0 &~~ \frac{\beta}{2\sqrt{3}}+\frac{\alpha}{\sqrt{2}}=0.7441 &a_0^+(1450) \rightarrow K^+\bar{K}^0 & -\frac{\alpha}{2\sqrt{3}}+\frac{\beta}{\sqrt{2}}=0.1722 \\
%K_0^{*+}(700) \rightarrow \pi^+K^0 & \frac{\beta}{2\sqrt{3}}+\frac{\alpha}{\sqrt{2}}=0.7430 & K_0^{*+}(1430)
%\rightarrow \pi^+ K^0 & -\frac{\alpha}{2\sqrt{3}}+\frac{\beta}{\sqrt{2}}=0.1770 \\
K_0^{*+}(700) \rightarrow \pi^0K^+ &~~ \frac{\beta}{2\sqrt{6}}+\frac{\alpha}{2}=0.5253 &
K_0^{*+}(1430) \rightarrow \pi^0 K^+ & -\frac{\alpha}{2\sqrt{6}}+\frac{\beta}{2}=0.1251 \\
f_0(500)\rightarrow \pi^0 \pi^0 &-\frac{a}{2}\left[\frac{\beta}{2\sqrt{3}}+\frac{\alpha}{\sqrt{2}}\right]=-0.3310 &  f_0(1370) \rightarrow \pi^0 \pi^0 & -\frac{a}{2}\left[-\frac{\alpha}{2\sqrt{3}}+\frac{\beta}{\sqrt{2}}\right]=-0.0785 \\
f_0(980)\rightarrow \pi^0 \pi^0 &~~\frac{b}{2}\left[\frac{\beta}{2\sqrt{3}}+\frac{\alpha}{\sqrt{2}}\right]=-0.1690 &  f_0(1500) \rightarrow \pi^0 \pi^0 & ~~\frac{b}{2}\left[-\frac{\alpha}{2\sqrt{3}}+\frac{\beta}{\sqrt{2}}\right]=-0.0394
\\
f_0(980) \rightarrow K^0\bar{K}^0 & -\frac{a}{2}\left[\frac{\beta}{2\sqrt{6}}+\frac{\alpha}{2}\right]=-0.4685  & f_0(1500) \rightarrow K^0\bar{K}^0 & -\frac{a}{2}\left[-\frac{\alpha}{2\sqrt{6}}+\frac{\beta}{2}\right]=-0.1093
\end{array}
\label{coupling}\
\end{eqnarray}
\end{widetext}
The flavor mixing parameters $a,b$ in the isoscalar resonances are
$a=0.8908$, $b=-0.4543$ fixed from the realistic case with fitting (RCF) in Ref.~\cite{Kim:2017yvd}.
Note also that $\alpha$, $\beta$ are slightly different depending on the isospin channel as given in Table~\ref{parameters}.
Some other channels related by isospin symmetry like $f_0(980)\rightarrow \pi^+ \pi^-$, $f_0(1500)\rightarrow \pi^+ \pi^-$, have not been listed here
because their information is not practically needed if our analysis is limited to the ratios
of the light nonet couplings over the heavy nonet couplings (see Sec.\ref{comparison}).

As advertised, the tetraquark mixing model indeed provides the anticipated feature that the coupling strengths are
enhanced in the light nonet and suppressed in the heavy nonet.
According to Eq.~(\ref{coupling}), the ratio of the light nonet couplings over the heavy nonet couplings
is around $G/G^\prime = 4.2 \text{--} 4.3$ depending on the channels so the relative enhancement factor is huge.
This fact is at least qualitatively consistent with the claims made in Sec.~\ref{implication} after the analysis of the experimental
partial widths.

%%%%%%%%%%%%%%%%%%%%%%%%%%%%%%%%%%%%%%%%%%%%%%%%%%%%%%%%%%%%%%%%%%%%%%%%%%%%%%%%%%%%%%%%%%%%%%
\section{Comparison of the experimental widths with theoretical predictions}
\label{comparison}
%%%%%%%%%%%%%%%%%%%%%%%%%%%%%%%%%%%%%%%%%%%%%%%%%%%%%%%%%%%%%%%%%%%%%%%%%%%%%%%%%%%%%%%%%%%%%%

So far, we have demonstrated
that the huge enhancement of the light nonet couplings relative to the heavy nonet can be seen
from the analysis of the current experimental partial widths from PDG~\cite{PDG22},
and claimed that the tetraquark mixing model can provide a dynamical reason for it at least in a qualitative level.
Then our next question is how good the agreement is {\it quantitatively}.
To discuss this, we calculate the theoretical partial widths
using the coupling strengths given in Eq.~(\ref{coupling}),
and compare them to the experimental partial widths in Table~\ref{partial width}.

Here we basically follow the formalism given in Ref.~\cite{Kim:2017yur}.
For the decay process, $a_0^+(980) \rightarrow \pi^+ \eta$, for example,
we take an effective Lagrangian involving derivatives~\cite{Black:1999yz},
\begin{eqnarray}
{\cal L}[a_0(980)\pi^+\eta]=(0.6076)~ g~ \partial_\mu \eta \partial^\mu \pi^+ a_0^+ (980)\ ,
\end{eqnarray}
with the overall constant $g$ and the coupling strength, $0.6067$, taken from Eq.~(\ref{coupling}) for this channel.
Using this, we then calculate the partial width with the $a_0(980)$ mass fixed to its central value, $M_c$,
\begin{eqnarray}
\Gamma [a_0\rightarrow \pi^+\eta]=\frac{(0.6076)^2 g^2 p}{32\pi {M_c^2}} (M_c^2-m_\pi^2-m_\eta^2)^2 .
\label{a0width}
\end{eqnarray}

But, since the $a_0(980)$ has a full decay width, its mass
should be distributed around the central value broadened by the full width.
This mass broadening can be implemented in the calculation of the partial decay width by
taking an average of the equation like Eq.~(\ref{a0width}) with respect to the mass distribution.
As discussed in Ref.~\cite{Kim:2017yur}, we take a mass distribution with the exponential type,
\begin{eqnarray}
f(M)\sim e^{-(M-M_c)^2 [4\!\ln2/\Gamma^2_{tot}]}\label{distribution}\ ,
\end{eqnarray}
where $\Gamma_{tot}$ is the total decay width of the resonance of concern, $a_0(980)$.
We then combine this with the decay width like Eq.~(\ref{a0width}) to obtain
the partial width averaged over the mass distribution,
\begin{eqnarray}
\langle \Gamma (M_c, \Gamma_{tot}) \rangle = \frac{\int^\infty_{m_{\pi}+m_{\eta}} \Gamma (M) f(M) dM}{\int^\infty_{m_{\pi}+m_{\eta}} f(M) dM}\ ,
\label{width av}
\end{eqnarray}
with respect to the inputs, $M_c$, $\Gamma_{tot}$.

Using this type of formula, Eq.~(\ref{width av}), we calculate theoretical partial widths for all the decay channels listed
in Eq.~(\ref{coupling}).
For $a_0(980)$, $f_0(980)$, their total widths are ranging from $50\text{--}100$ MeV, $10\text{--}100$ MeV respectively in Table~\ref{two nonets}.
But their partial widths listed in Table~\ref{partial width}, even though they do not represent all the possible partial decay modes,
are already greater than $60$ MeV when they are summed.  To choose the total width bigger than $60$ MeV,
we take the total widths of $a_0(980)$, $f_0(980)$ to be $100$ MeV in our theoretical calculation. But the smaller total width
like $60$ MeV does not change much the ratios reported below.

To avoid other complications, we consider in this work the ratios of the
light-nonet partial widths over the corresponding heavy-nonet partial widths.
These ratios, first of all, eliminate the dependence on the overall unknown constant, $g$, in the couplings and, secondly, they are
free from the $\eta-\eta^\prime$ mixing especially for the decay involving $\eta$,
and finally, they make the decay modes presented in Eq.~(\ref{coupling}) enough for comparison with the experimental ratios
even if they have additional isospin channels that are not listed in Eq.~(\ref{coupling}).

Here we present the ratios of the partial decay widths theoretically calculated using Eq.~(\ref{width av}) and
the ratios of the experimental partial widths extracted from Table~\ref{partial width}.
\begin{eqnarray}
&                            &~~~~~~~~~~~~~~~~~~~~~~~~~~~~~~\text{\underline{Theory}} ~~~~~~~~~~~\text{\underline{Expt.}} \nonumber \\
& &\frac{\!\!\!\Gamma [a_0(980)\rightarrow \pi \eta]}{\Gamma [a_0(1450)\rightarrow \pi \eta]} ~~:~~~ 2.54~~~~~~~~~~~2.93 \text{--} 3.9 \label{rat1}\ \\
\nonumber \\
& &\frac{\!\!\!\Gamma [a_0(980)\rightarrow K\bar{K}]}{\Gamma [a_0(1450)\rightarrow K\bar{K}]}  :~~~ 0.89~~~~~~~~~~0.59 \text{--} 0.79\label{rat2}\ \\
\nonumber \\
& &\frac{\!\!\!\Gamma [K_0^*(700) \rightarrow \pi K]}{\Gamma [K_0^*(1430) \rightarrow \pi K]}  :~~~ 2.58~~~~~~~~~~~~~1.86\label{rat3}\ \\
\nonumber \\
& &\frac{\!\!\!\Gamma [f_0(980)\rightarrow \pi \pi]}{\Gamma [f_0(1500)\rightarrow \pi \pi]}    ~~:~~~ 4.90~~~~~~~~~~~~~1.31\label{rat4}\ \\
\nonumber \\
& &\frac{\!\!\!\Gamma [f_0(980)\rightarrow K\bar{K}]}{\Gamma [f_0(1500)\rightarrow K\bar{K}]}  :~~~ 1.73~~~~~~~~~~1.00 \text{--} 4.86\label{rat5}\
\end{eqnarray}
Note, the $a_0(980)$ mass, $M_c=980$ MeV, is just below the $K\bar{K}$ threshold $\sim 990$ MeV.
So the decay like $a_0(980)\rightarrow K\bar{K}$ in the theoretical calculation is mostly prohibited by this kinematical cutoff, and it happens
only when the mass broadening through the total width is taken into account.
The similar discussion can be made for $f_0(980) \rightarrow K\bar{K}$.
Thus, the averaging step like Eq.~(\ref{width av}) is clearly needed to have nonzero partial widths especially for $a_0(980)\rightarrow K\bar{K}$,
$f_0(980) \rightarrow K\bar{K}$.

But because of this, the partial widths for $a_0(980)\rightarrow K\bar{K}$, $f_0(980)\rightarrow K\bar{K}$
relative to the corresponding heavy nonet widths should be
suppressed further by the kinematical cutoff in the mass distribution
in addition to the suppression due to the mass gap between the initial and final decay products.
Even so, the theoretical ratios,
$\Gamma [a_0(980)\rightarrow K\bar{K}]/\Gamma [a_0(1450)\rightarrow K\bar{K}]=0.89$ in Eq.~(\ref{rat2}),
$\Gamma [f_0(980)\rightarrow K\bar{K}]/\Gamma [f_0(1500)\rightarrow K\bar{K}]=1.73$ in Eq.~(\ref{rat5}),
clearly show that the two modes in the light nonet are not suppressed at all
and, as mentioned already, this obviously has to come from the strong enhancement in the couplings of the light nonet.

What is interesting is the fact that such theoretical ratios
agree relatively well with the corresponding experimental ratios in Eqs.~(\ref{rat2}), (\ref{rat5}),
although the agreement in Eq.~(\ref{rat5}) has to be taken with some cautions because of the large uncertainty in
the experimental ratio.
We also see in Eq.~(\ref{rat1}) that the theoretical ratio of $\Gamma [a_0(980)\rightarrow \pi \eta]/\Gamma [a_0(1450)\rightarrow \pi \eta]\sim 2.54$
is less than the experimental ratio range, $2.93\text{--}3.8$, still showing an agreement with a reasonable degree.
Therefore, we may conclude that Eqs.~(\ref{rat1}), (\ref{rat2}), (\ref{rat5}) support the tetraquark mixing model
even in a quantitative level.

But for the other ratios in Eqs.~(\ref{rat3}), (\ref{rat4}), the agreement is not so good as
the theoretical ratios are much larger than the experimental ratios.
For $\Gamma [f_0(980)\rightarrow \pi\pi]/\Gamma [f_0(1500)\rightarrow \pi\pi]$, the theoretical ratio of $4.90$ is
much higher than the experimental ratio $1.31$.
For $\Gamma [K_0^*(700) \rightarrow \pi K]/\Gamma [K_0^*(1430) \rightarrow \pi K]$,
its theoretical ratio is $2.58$, about $40\%$ larger than the experimental ratio.
Even though the experimental ratios do not quantitatively agree with the theoretical ones,
these two still support that the coupling strengths of the light nonet are enhanced over those of the heavy nonet.
Nevertheless, we can look for possible improvements both experimentally and theoretically.

In calculating the theoretical ratio of $\Gamma [K_0^*(700) \rightarrow \pi K]/\Gamma [K_0^*(1430) \rightarrow \pi K]=2.58$,
we have used, as the inputs, the experimental values of the central mass, $M[K_0^*(700)]=845$ MeV, and the total width,
$\Gamma_{tot}=468$ MeV, obtained from the current PDG~\cite{PDG22}.
But their values were $M[K_0^*(700)]=682$ MeV, $\Gamma_{tot}=547$ MeV in the 2016 version of PDG~\cite{PDG16}.
If these old values were used instead,
the calculated ratio becomes~\footnote{The same value was reported in Ref.~\cite{Kim:2017yur} but at that time the experimental data for the partial width
were not available to calculate the corresponding experimental ratio.} $1.76$, which is much lower than the current theoretical ratio.
This reduces the disagreement with the experimental ratio around $5\%$.
Similarly, this type of variation due to experimental ambiguity can be expected also for Eq.~(\ref{rat4}) and others, Eqs.~(\ref{rat1}),(\ref{rat1}),(\ref{rat5}).
Therefore, based on this examination, our quantitative comparison is still variable depending on the update on PDG data.

Alternatively, one can look for improvements from theoretical side.
One possibility is that the tetraquark mixing model may not fully describe the two nonets in PDG.
In fact, there are various other models for the two nonets in the literature, and one may wish to find a
better description by combining our model with others.
For example, Refs.~\cite{Black:1998wt,Black:1999yz} propose a different model where
the $P$-wave $q\bar{q}$ mix with the four-quark
$qq\bar{q}\bar{q}$ nonet~\cite{Black:1998wt,Black:1999yz}. If this picture is somehow combined
with our tetraquark mixing model, this can modify the coupling strengths given in
Eq.~(\ref{coupling}) and change the partial decay widths accordingly.
In the quark level, however, the two-quark states, $q\bar{q}$, do not mix with the four-quark states
through the color-spin interaction in a simple approach.  Thus, even if one can come up with
a certain framework that combines this model with our model, the modification due
to this is expected to be small.
Or one can think about combining our mixing model with the hadronic molecular picture~\cite{Janssen:1994wn,Branz:2007xp,Branz:2008ha} or other pictures
as in Refs.~\cite{Dudek:2016cru,Tornqvist:1995kr, vanBeveren:1986ea,Maiani:2006rq}
with a hope that certain improvement is achieved.
However, it seems that all these scenarios are not easy to proceed at this point.

Nevertheless, what we want to stress is that any
ratio in Eqs.~(\ref{rat1}), $\cdot\cdot\cdot$, (\ref{rat5}) does not undermine the
qualitative conclusion made in Sec.~\ref{couplings from TMM}, namely, the conclusion that
the coupling strengths of the light nonet are enhanced over those of the heavy nonet.
Even though certain improvement might be expected in future,
it is clear that the tetraquark mixing model in current form already plays a unique role in explaining the
interesting aspect seen in the coupling strengths of the two nonets.
Therefore, the tetraquark mixing model can be a strong candidate as a possible structure for the light nonet and the heavy nonet
in PDG.

\section{Summary}
\label{sec:summary}

In this work, we have investigated some evidences from the partial decay widths
that support tetraquark mixing framework for the two nonets in PDG, $a_0 (980)$, $K_0^* (800)$, $f_0 (500)$, $f_0 (980)$ in the light nonet,
$a_0 (1450)$, $K_0^* (1430)$, $f_0 (1370)$, $f_0 (1500)$ in the heavy nonet.
In particular, we have analyzed the partial decay widths of the two nonets
from the PDG data and found that the partial widths from the light nonet are generally larger than those from the heavy nonet.
This peculiar aspect is opposite to the natural expectation that
the heavy nonet may have larger widths than the light nonet due to the larger mass gap between the initial and final decay products.
To explain this opposite trend,
we have come up with the fact that the coupling strength in each channel must be enhanced in the light nonet
and suppressed in the heavy nonet.
This phenomenological aspect in the couplings in fact can be qualitatively generated from
the tetraquark mixing model that has been proposed as a possible structure for the two nonets in PDG.
In this model, the two nonets
are generated by the mixing of two tetraquark configurations and,
as a result, the coupling strengths are enhanced in the light nonet and suppressed in the heavy nonet.

This qualitative agreement has been tested further by calculating
explicitly the partial decay widths using the coupling strengths obtained from the tetraquark mixing model.
We then compare the ratios of the partial widths in each isospin channel with the experimental ratios.
We have reported that some ratios compare relatively well and some other ratios do not, and that the comparison
could depend on the update in the PDG data. But, at least, none of them undermine the general conclusion
that the coupling strengths in the light nonet are enhanced over those in the heavy nonet.
We believe that this clearly supports the tetraquark mixing model
as a possible structure for the two nonets in PDG.

\acknowledgments

\newblock
This work was supported by the National Research Foundation of Korea(NRF) grant funded by the
Korea government(MSIT) (No. NRF-2018R1A2B6002432,
No. NRF-2018R1A5A1025563).

\end{document}